\documentclass[cameraready]{Interspeech}

\title{SARA: A Dual-Stream VAE for High-Fidelity Speech Generation via Integrating Semantic and Acoustic Representations\thanks{$^{*}$ Work done during an internship at DiDi Global Inc.}\thanks{$^\dagger$ Corresponding author}}

\author[affiliation={1}]{Peijie}{Chen$^{*,\,}$}
\author[affiliation={2}]{Wenhao}{Guan}
\author[affiliation={1}]{Weijie}{Wu}
\author[affiliation={1}]{Kadi}{Wang}
\author[affiliation={3}]{Daiyu}{Huang}
\author[affiliation={3}]{Zhuanling}{Zha}
\author[affiliation={3}]{Junbo}{Li}
\author[affiliation={3}]{Jun}{Fang}
\author[affiliation={1}]{Qingyang}{Hong$^{\dagger,\,}$}
\author[affiliation={2}]{Lin}{Li}

\address{
    $^1$ School of Informatics, Xiamen University, China \\
    $^2$ School of Electronic Science and Engineering, Xiamen University, China \\
    $^3$ DiDi Global Inc., Beijing, China
}

\email{peijiechen@stu.xmu.edu.cn}


\usepackage{comment}
\usepackage{amsmath}
\usepackage{multirow}
\usepackage{arydshln}
\usepackage{booktabs}
\usepackage{subcaption}
\usepackage{hyperref}

\begin{document}

\maketitle
\begin{abstract}
Zero-shot text-to-speech (TTS) relies on robust speech representations. However, current speech tokenizers face a fundamental trade-off: acoustic codecs preserve high-fidelity audio but lack linguistic constraints, causing content errors during generation, whereas semantic tokens from self-supervised learning (SSL) models ensure precise text alignment but discard some acoustic information. To bridge this gap, we propose SARA, a dual-stream VAE that directly fuses a frozen SSL semantic anchor with a dedicated residual acoustic encoder. This effectively mitigates the dilemma, creating an efficient and compact latent space without relying on complex regularizers. SARA achieves superior reconstruction quality over strong baselines. Furthermore, in downstream zero-shot TTS tasks, it yields highly natural and expressive synthesis quality, and maintains robust generation performance even under accelerated inference, offering a favorable trade-off between synthesis speed and computational cost.
\end{abstract}

\section{Introduction}
    Recent advancements in large-scale speech generation models have significantly propelled the capabilities of zero-shot Text-to-Speech (TTS) systems\cite{du2024cosyvoice,du2024cosyvoice1,du2025cosyvoice,zhou2025indextts2}, enabling the synthesis of highly natural and speaker-adaptive voices from minimal prompt data. A fundamental component of these modern architectures is the speech tokenizer \cite{zhang2023speechtokenizer} or neural speech codec\cite{kumar2023high,defossez2022highfi,xin2024bigcodec,chen2025ds}, which compresses continuous, high-dimensional audio waveforms into low-dimensional latent representations. The quality and nature of this latent space dictate not only the theoretical upper bound of the synthesized audio quality but also the stability and convergence speed of the downstream autoregressive\cite{jia2025ditar,zhou2025voxcpm} or non-autoregressive\cite{chen2025f5,eskimez2024e2,lee2024ditto} generative models.

    However, designing an optimal continuous or discrete representation remains a formidable challenge due to the inherent trade-off between reconstruction fidelity and generative controllability. Traditional acoustic codecs excel at reconstructing fine-grained physical details, such as high-frequency harmonics and subtle environmental noises, but their latent spaces lack explicit linguistic constraints. Consequently, when utilized in downstream zero-shot TTS, these representations often fail to provide strong semantic guidance, leading to content inaccuracies and elevated Word Error Rates (WER). Conversely, purely semantic representations derived from pre-trained self-supervised learning (SSL) models (e.g., HuBERT\cite{hsu2021hubert}, WavLM\cite{chen2022wavlm}, W2v-BERT\cite{chung2021w2v}) or ASR models offer exceptional content alignment. Yet, they inherently discard crucial acoustic characteristics, including speaker identity, emotion, and prosody, resulting in synthesized speech with low speaker similarity and degraded perceptual quality. Existing attempts to bridge this gap, such as introducing complex semantic regularization losses \cite{niu2025semantic,leng2025repa} during VAE training, are often indirect and difficult to balance.
    
    To address this dilemma, we propose SARA (Semantic-Acoustic Residual Autoencoder), a novel dual-stream VAE framework designed to seamlessly integrate high-level linguistic content with fine-grained acoustic details. Rather than relying on complex regularization terms, SARA adopts a structural approach to complementary representation integration. It pairs a frozen pre-trained SSL model acting as a stable semantic anchor with a dedicated residual acoustic encoder tasked with capturing the rich timbre omitted by the semantic branch. By fusing these inherently aligned streams, SARA creates a compact and efficient information bottleneck. The entire framework is optimized via a VAE paradigm, ensuring a well-regularized latent space for downstream generative modeling and exceptional perceptual quality for waveform reconstruction.
    
    The main contributions are summarized as follows:
        \begin{itemize}
            \item We introduce SARA, a dual-stream variational autoencoder that effectively mitigates the semantic-acoustic trade-off via a direct and highly compressed feature integration mechanism, eliminating the need for complex training regularizers.
            \item Extensive evaluations demonstrate that SARA achieves excellent speech reconstruction fidelity compared to strong VAE baselines, all while maintaining an extremely compact latent space.
            \item When in a zero-shot TTS framework, SARA significantly improves both content accuracy and speaker similarity, validating the powerful generalization capabilities of its learned representations.
        \end{itemize}

    Audio samples of the reconstructed speech and downstream zero-shot synthesized speech from various models are available at https://pppjchen.github.io/SARA.
\section{Method}
\begin{figure*}[t]
  \centering
  \includegraphics[width=0.85\linewidth]{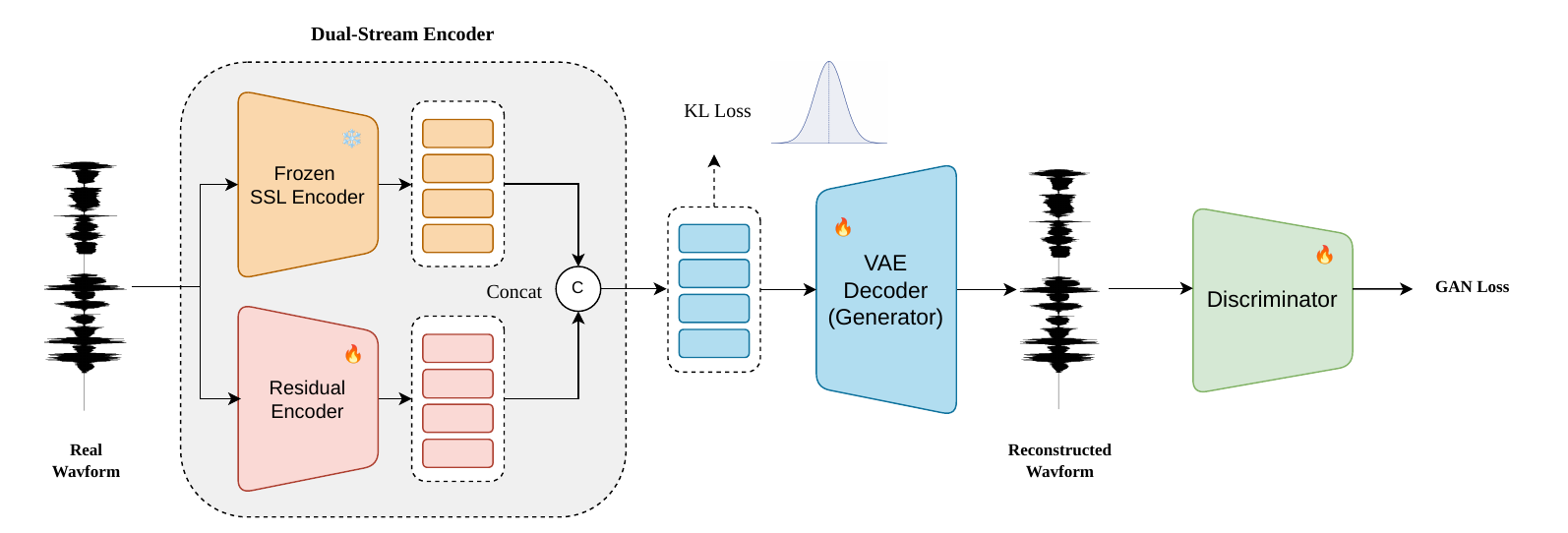}
  \caption{Overall model structure. The frozen SSL Encoder extracts semantic representations, while the Residual Acoustic Encoder captures fine-grained acoustic details.}
  \label{fig:model}
\end{figure*}

As illustrated in Fig.~\ref{fig:model}, we propose SARA, a dual-stream variational autoencoder (VAE) that effectively mitigates the trade-off between reconstruction fidelity and generative quality.

Section \ref{section:vae} first formulates the basic VAE and its training objectives. Section \ref{section:semantic_injection} then explains our semantic injection strategy. Finally, Section \ref{section:architecture} details the proposed network architecture.

\subsection{Variational Autoencoder}
\label{section:vae}
Our framework leverages a Variational Autoencoder (VAE) to extract continuous latent representations from raw speech waveforms. The architecture comprises an encoder that projects the input signal $x$ into a latent variable $z$, and a decoder that reconstructs the waveform $\hat{x}$ from this latent space. Optimization proceeds by maximizing the Evidence Lower Bound (ELBO):

\begin{equation}
    \label{eq:elbo}
    \scalebox{0.95}{$\displaystyle
    \text{ELBO} := \mathbb{E}_{q_\phi(z|x)}\left[\log p_\psi(x|z)\right] - \mathcal{D}_\mathrm{KL}\left[q_\phi(z|x) \parallel p(z)\right]
    $}
\end{equation}

In this context, $q_\phi(z|x)$ represents the variational posterior approximating the true distribution, while $p_\psi(x|z)$ defines the likelihood of the generated output. The Kullback-Leibler (KL) divergence term constrains the latent distribution to align with the standard Gaussian prior $p(z)=\mathcal{N}(0, I)$.

To operationalize this objective, we minimize the negative ELBO, formulated as a weighted combination of reconstruction and regularization terms. For the reconstruction loss $\mathcal{L}_{\text{recon}}$, we adopt the multi-scale mel-spectrogram loss consistent with DAC \cite{kumar2023high}. To further augment perceptual quality, we integrate adversarial training employing a multi-period discriminator \cite{kong2020hifi} and a multi-band \cite{kumar2023high}, multi-scale STFT discriminator, which is optimized by an L1 feature-matching loss ($\mathcal{L}_{\text{feat}}$). 

The complete VAE training objective is expressed as follows:

\begin{equation}
\mathcal{L}_{\text{VAE}}
= \lambda_{\text{recon}}\,\mathcal{L}_{\text{recon}}
+ \lambda_{\text{KL}}\,\mathcal{L}_{\text{KL}}
+ \lambda_{\text{adv}}\,\mathcal{L}_{\text{adv}}
+ \lambda_{\text{feat}}\,\mathcal{L}_{\text{feat}}.
\end{equation}

\subsection{Semantic Injection}
\label{section:semantic_injection}
Striking a balance between reconstruction fidelity and generative controllability constitutes a core obstacle in developing discrete neural speech codecs\cite{zhang2023speechtokenizer}. Prior research indicates that incorporating guidance from pre-trained SSL models (e.g., HuBERT\cite{hsu2021hubert}, WavLM\cite{chen2022wavlm}, W2v-BERT\cite{chung2021w2v}) offers a viable pathway. By leveraging the rich semantic representations embedded within these models, codecs receive more effective supervision, thereby enhancing the convergence and performance of downstream auto-regressive systems\cite{li2025dualcodec}.

Recently, Semantic-VAE\cite{niu2025semantic} introduced a semantic regularization loss into the VAE training framework to improve downstream TTS performance. In contrast, we propose an architectural innovation that directly integrates a frozen SSL model into the VAE encoder, paired with a residual acoustic encoder, eliminating the need for additional regularization losses. The frozen SSL branch acts as a stable content anchor, supplying robust semantic representations. Concurrently, the residual acoustic branch captures fine-grained acoustic details omitted by the SSL model. This complementary representation learning allows us to directly navigate the reconstruction-generation dilemma: the semantic branch ensures content consistency for generation, while the residual branch preserves high-fidelity details. Consequently, SARA yields latent representations that are both semantically meaningful for downstream modeling and acoustically rich for high-quality speech synthesis.
\begin{table*}[t]
\centering
\caption{Performance comparison of the proposed SARA framework. Zero-shot TTS performance is evaluated on the LibriSpeech-PC test-clean set using F5-TTS\cite{chen2025f5} as the generation backbone. * means the results are obtained from the paper.}
\label{tab:main_results}

    \begin{tabular}{lccccccc}
    \toprule
    \textbf{Model} & \textbf{\#Param.} & \textbf{Sample Rate} & \textbf{Frame/s} & \textbf{WER(\%)}$\downarrow$ & \textbf{SIM}$\uparrow$ & \textbf{CMOS}$\uparrow$ & \textbf{SMOS}$\uparrow$ \\
    \midrule
    GT & - & - & - &2.23 & 0.69 & +0.12 & 3.92 \\
    Vocoder Resynthesized & - & 24k & - & 2.32 & 0.66 & +0.10 & 3.91 \\
    \midrule
    Cosyvoice & 300M & 24k & - & 3.59 & 0.66 & -0.14 & 3.95 \\
    E2 TTS & 333M & 24k & -  & 2.95 & 0.69 & -0.08 & 3.98 \\
    F5-TTS & 336M & 24k & -  & 2.42 & 0.66 & -0.06 & 3.99 \\
    \midrule
    F5-TTS-Small & 159M & 24k  & 93.75 & 2.23 & 0.60 & -0.10 & 3.85 \\

    $\quad$ + Semantic-VAE* & 159M  & 16k & 40 & 1.95 & 0.64 & - & - \\
    $\quad$ + SARA (Ours) & 159M  & 24k & 50 & \textbf{1.79} & 0.63 &  -0.03 & 3.89 \\
    \midrule
    \textit{Scaling Up} & & & &  & \\
    F5-TTS-Base + SARA & 336M & 24k & 50 & \textbf{1.74} & \textbf{0.655} & 0.00 & 3.90\\
    \bottomrule
    \end{tabular}
\end{table*}

\subsection{Model Architecture}
\label{section:architecture}
The proposed SARA framework adopts a dual-stream encoder architecture designed to capture complementary speech information. The system maps a 24 kHz input waveform $x$ to a 64-dimensional latent representation $z$ at a frame rate of 50 Hz, which is subsequently reconstructed by a high-fidelity decoder.

\subsubsection{Acoustic and Semantic Encoders}
Following the design principles of BigCodec\cite{xin2024bigcodec}, the residual acoustic encoder is built upon a series of residual Convolutional Neural Network (CNN) blocks. Each block incorporates Snake activation functions and multiple convolutional layers with varying dilation rates to model multi-scale sequential patterns effectively. To capture long-range temporal dependencies, the CNN output is processed by a two-layer unidirectional Long Short-Term Memory (LSTM) network.
In parallel, we employ W2v-BERT 2.0\cite{chung2021w2v} as a frozen SSL encoder to extract robust semantic representations. These two independent branches are designed to comprehensively process the input waveform before feature integration.

The residual acoustic encoder pipeline achieves a cumulative downsampling factor of 480 through five successive modules with striding factors of [2, 3, 4, 4, 5]. This configuration effectively compresses the 24 kHz audio signal into a compact 50 Hz acoustic latent stream ($z_{\text{ac}}$). 

Conveniently, the frozen W2v-BERT 2.0\cite{chung2021w2v} encoder inherently extracts semantic representations ($z_{\text{sem}}$) at an identical frame rate of 50 Hz. Leveraging this temporal alignment, we directly concatenate the semantic and acoustic streams along the channel dimension. The fused feature map is subsequently passed through a linear projection layer to map it to the target fixed dimension, yielding the final 64-dimensional latent representation $z$. This straightforward yet highly effective fusion mechanism maintains an efficient information bottleneck for downstream tasks, seamlessly integrating high-level linguistic content with fine-grained acoustic details.

\subsubsection{Decoder}
The decoder is based on the HiFi-GAN\cite{kong2020hifi} architecture, utilizing multi-receptive field fusion to synthesize high-fidelity waveforms from the integrated latent representations. The entire framework is optimized using an adversarial training paradigm to ensure perceptual naturalness and reconstruction accuracy.

\section{Experimental Setup}

\subsection{Datasets and Metrics}
We utilize a large-scale corpus consisting of LibriTTS\cite{zen2019libritts} and LibriHeavy\cite{kang2024libriheavy} to train the proposed SARA model. LibriHeavy provides approximately 50,000 hours of audiobook speech (originally at 16~kHz), while LibriTTS contributes 585 hours of high-quality, multi-speaker speech at 24~kHz. To unify the sampling rate across the corpus, all training data is resampled to 24~kHz, totaling over 50,000 hours of diverse audio. For the downstream generative task, the zero-shot TTS model is trained exclusively on the LibriHeavy subset.

To evaluate the efficacy of SARA in balancing reconstruction fidelity with generative utility, we conduct assessments across two primary benchmarks:

\begin{itemize}
    \item \textbf{Speech Reconstruction and Latent Utility}: We evaluate reconstruction performance on the test-clean split of LibriSpeech\cite{panayotov2015librispeech}. Objective metrics including PESQ\cite{rix2001perceptual}, STOI, and UTMOS\cite{saeki2022utmos} are employed to quantify perceptual quality and intelligibility. Furthermore, to verify the semantic robustness and speaker-independent nature of the latent space, we perform cross-sentence generation. This is assessed via WER for content accuracy, SIM for voice consistency, and UTMOS for overall naturalness.

    \item \textbf{Downstream Zero-Shot TTS}: We assess the zero-shot synthesis performance on the LibriSpeech-PC test-clean set\cite{panayotov2015librispeech,chen2025f5,meister2023librispeech}. For objective evaluations, we utilize the Whisper-large-v3\cite{radford2023robust} model to compute the WER. To measure SIM, we calculate the cosine similarity between the generated speech and the reference audio using speaker embeddings extracted from a pre-trained WavLM-TDCNN model. Furthermore, we conduct subjective evaluations across two primary dimensions: CMOS, to assess overall audio quality, clarity, naturalness, and high-fidelity details; and SMOS, to evaluate speaker similarity in terms of timbre reconstruction and prosodic patterns.
\end{itemize}
    
\subsection{Training Details}
\subsubsection{VAE Setup}
SARA is optimized for 200k iterations using a global batch size of 256. For a fair comparison, all ablation studies share the same training configuration. Audio samples are segmented into 1-second clips and resampled to 24kHz, resulting in approximately 256 seconds of audio per batch. We employ the AdamW optimizer with an initial learning rate of \(1 \times 10^{-4}\). A linear warm-up schedule is applied to both the learning rate and the $\lambda_\text{KL}$ coefficient for the first 10,000 updates, followed by an exponential decay with a factor of $\gamma = 0.9999996$. The loss weighting coefficients are empirically set as: $\lambda_\text{KL}=0.01$, $\lambda_\text{adv}=1$, $\lambda_\text{feat}=1$, and $\lambda_\text{recon}=15$.

\subsubsection{Downstream Zero-shot TTS Model Setup}
For the downstream zero-shot TTS task, we adopt F5-TTS as the generation backbone and follow its original training configuration. Specifically, we substitute the conventional mel-spectrograms with the latent representations extracted from the SARA dual-stream encoder. The model is optimized using the AdamW optimizer with a peak learning rate of \(7.5 \times 10^{-5}\), incorporating a 20,000-step warm-up phase followed by linear decay. During inference, we maintain consistency with the F5-TTS protocol, employing a sway sampling strategy and an Euler ODE solver for high-quality speech synthesis.

\subsection{Result}
\subsubsection{Downstream Zero-shot TTS Results}
    The downstream zero-shot TTS performance is summarized in Table \ref{tab:main_results}. A key strength of the SARA framework lies in its exceptional content preservation capabilities. When integrated into the F5-TTS-Small, SARA achieves an impressive WER of 1.79, significantly outperforming the vanilla baseline. Remarkably, this compact configuration surpasses parameter-heavy baselines, including Cosyvoice\cite{du2024cosyvoice1}, E2 TTS\cite{eskimez2024e2}, and the standard F5-TTS. Beyond content accuracy, SARA maintains robust acoustic fidelity, surpassing the vanilla model in objective speaker similarity. Notably, SARA circumvents the 16 kHz bandwidth limitation of the Semantic-VAE baseline, supporting high-fidelity synthesis while simultaneously delivering lower WER.

    Crucially, the proposed representations exhibit strong scalability. When applied to the F5-TTS (Base), SARA further drives the WER down to 1.74 and elevates the SIM to 0.655. These results conclusively validate SARA's efficacy in successfully navigating the trade-off between semantic controllability and high-fidelity acoustic generation.

\subsubsection{Reconstruction Results}
    Table \ref{tab:recon_results} presents the objective reconstruction results on the LibriSpeech test-clean set. Despite operating at a compact latent, the proposed SARA framework also demonstrates exceptional reconstruction fidelity. SARA achieves the highest PESQ and STOI, significantly outperforming the high-bandwidth Vocos\cite{siuzdak2023vocos} baseline and the Semantic-VAE. Notably, the substantial improvement in PESQ over the Vanilla VAE directly validates that the dual-stream architecture, pairing a semantic anchor with a residual acoustic encoder, effectively recovers fine-grained acoustic details. Furthermore, while Semantic-VAE yields a higher UTMOS score, SARA maintains a highly competitive score of 4.100, which effectively surpasses the GT, confirming its robust perceptual naturalness.
    
\begin{table}[h] 
\centering
\caption{Speech reconstruction performance on the LibriSpeech test-clean set.}
\label{tab:recon_results}
\resizebox{\linewidth}{!}{
\begin{tabular}{lccccc}
\toprule
\textbf{Model} & \textbf{Frame/s} & \textbf{Dim} & \textbf{PESQ}$\uparrow$ & \textbf{STOI}$\uparrow$ & \textbf{UTMOS}$\uparrow$ \\
\midrule
GT & - & - & - & - & 4.086 \\
Vocos & 93.75 & 100 & 3.605 & 0.977 & 3.625 \\
Semantic-VAE & 40 & 64 & 3.968 & 0.981 & \textbf{4.129} \\ 
Vanilla VAE & 50 & 64 & 4.076 & 0.983 & 4.095 \\
\midrule
\textbf{SARA (Ours)} & 50 & 64 & \textbf{4.389} & \textbf{0.993}& \textbf{4.100} \\
\bottomrule
\end{tabular}
}
\end{table}
\subsection{Ablation Study}
    We conduct ablation studies to evaluate the effectiveness of the proposed dual-stream architecture. Specifically, we evaluate the reconstruction capabilities of the VAE models on the LibriSpeech-PC test-clean set. To explicitly isolate the contribution of the dual-stream encoder, we compute not only standard acoustic quality metrics but also the WER and SIM against the ground truth reference audio.
\begin{table}[ht] 
\centering
\caption{Ablation study of speech reconstruction on the LibriSpeech-PC test-clean set.}
\label{tab:ablation}
\resizebox{\linewidth}{!}{
\begin{tabular}{lccccc}
\toprule
\textbf{Model} & \textbf{PESQ}$\uparrow$ & \textbf{STOI}$\uparrow$ & \textbf{UTMOS}$\uparrow$ & \textbf{WER(\%)}$\downarrow$ & \textbf{SIM}$\uparrow$\\
\midrule
GT & - & - & 4.097 & 2.23 &  0.690\\
SARA & \textbf{4.366} & \textbf{0.992} & \textbf{4.110} &  \textbf{2.32} & \textbf{0.685} \\
$\quad$- Res Encoder & 2.655 & 0.930 & 3.944 &  2.41 & 0.640 \\
$\quad$- SSL Encoder & 4.074 & 0.983 & 4.113 &  2.41 & 0.683 \\
\bottomrule
\end{tabular}
}
\end{table}

Table 3 presents an ablation study on the LibriSpeech-PC test-clean set to evaluate the individual contributions of the dual-stream components within the SARA framework. The removal of the residual acoustic encoder (denoted as "- Res Encoder") leads to a measurable decline in speaker consistency, with the SIM score decreasing from 0.685 to 0.640. Furthermore, objective reconstruction metrics (PESQ, STOI) experience noticeable drops. This indicates that the frozen SSL semantic branch, while rich in content representation, is insufficient on its own to reconstruct the original speaker's timbre and fine-grained acoustic details. Conversely, omitting the pre-trained SSL semantic encoder (denoted as "- SSL Encoder") effectively reduces the architecture to a Vanilla VAE and compromises the linguistic stability of the latent space. Consequently, content accuracy degrades, as evidenced by an increase in WER from 2.32 to 2.41.

Together, these findings support the design motivation of SARA: the semantic and acoustic branches capture complementary information. The dual-stream integration effectively balances semantic consistency for generation with the acoustic requirements for high-fidelity speech reconstruction.

\begin{table}[ht] 
\centering
\caption{Ablation study on flow matching inference steps for F5-TTS Small integrated with SARA on the LibriSpeech-PC test-clean set. The Real-Time Factor (RTF) is computed with the average of the inference time of the test-set on a single NVIDIA A100 80GB GPU. }
\label{tab:nfe_result}
\resizebox{\linewidth}{!}{
\begin{tabular}{lccccc}
\toprule
\textbf{Model} & \textbf{NFE} & \textbf{WER(\%)}$\downarrow$ & \textbf{SIM}$\uparrow$ & \textbf{RTF}$\downarrow$\\
\midrule
GT & - & 2.23 &  0.69 & -\\
\midrule
F5-TTS-Small & 8 &  3.51 & 0.58 & 0.061 \\
F5-TTS-Small & 32 &  2.23 & 0.60 & 0.115 \\
\midrule
F5-TTS-Small + SARA & 6 &  2.27 & 0.57 & 0.058\\
F5-TTS-Small + SARA & 8 &  1.82 & 0.62 &  0.079\\
F5-TTS-Small + SARA & 32 &  1.79 & 0.63 &  0.184\\
\bottomrule
\end{tabular}
}
\end{table}

In flow matching paradigms, the ability to maintain synthesis quality at lower NFE typically indicates a well-conditioned target latent space. To evaluate this, we analyze the impact of reduced inference steps in Table 4. Remarkably, decreasing the NFE from 32 to 8 yields only negligible shifts in both WER and SIM. This strong performance suggests that SARA's dual-stream architecture effectively regularizes the generative trajectory. The robust semantic anchors reduce the prediction ambiguity, allowing the flow model to converge rapidly and synthesize high-fidelity speech. Although the dual-stream encoder introduces extra inference cost, SARA with 8 steps (RTF 0.079) achieves comparable synthesis quality to the 32-step F5-TTS-small baseline (RTF 0.115) while offering a more favorable trade-off in generation speed.
\section{Conclusion}
We presented SARA, a novel dual-stream VAE designed to mitigate the gap between high-fidelity acoustic reconstruction and robust generative controllability. SARA seamlessly integrates high-level linguistic content and fine-grained acoustic details by pairing a frozen SSL branch with a residual acoustic encoder. Leveraging their temporal alignment, we established an efficient 50Hz, 64-dimensional information bottleneck. Experiments confirm that SARA surpasses robust baselines in speech reconstruction intelligibility and quality. Crucially, utilizing SARA's latent space within the F5-TTS framework yields substantial gains in zero-shot content accuracy and speaker similarity. Ultimately, SARA provides a versatile foundation for large-scale speech generation, with future efforts focused on multilingual scaling and integrating autoregressive models.

\section{Generative AI Use Disclosure}
During the preparation of this manuscript, the authors used generative AI tools to polish the English language, improve readability, and assist with \LaTeX\xspace formatting. These tools were not used to generate any scientific claims, experimental results, or significant parts of the manuscript.
\section{Acknowledgements}
This work was supported in part by the National Natural Science Foundation of China under Grants 62276220 and 62371407 and the Innovation of Policing Science and Technology, Fujian province (Grant number: 2024Y0068)

\bibliographystyle{IEEEtran}
\bibliography{mybib}

\end{document}